\documentclass[11pt]{article}
\usepackage{titlesec}
\titleformat{\paragraph}[runin]{\normalfont\itshape}{\theparagraph.}{.3em}{}[.]\titlespacing{\paragraph}{0pt}{1ex plus .1ex minus .2ex}{.5em}
\usepackage{amsmath,amssymb,bm}
\usepackage{mathabx}
\usepackage[T1]{fontenc}
\usepackage[utf8]{inputenc}
\usepackage{lmodern}
\usepackage{mathtools}
\usepackage{dsfont}
\pdfoutput=1
\usepackage[english]{babel}
\usepackage[letterpaper, hmargin=1in, top=1in, bottom=1.2in, footskip=0.6in]{geometry}
\usepackage{graphicx} 
\usepackage{booktabs} 
\usepackage{color}
\definecolor{aquamarine}{rgb}{0.5, 1.0, 0.83}
\definecolor{ao(english)}{rgb}{0.0, 0.5, 0.0}
\definecolor{armygreen}{rgb}{0.29, 0.33, 0.13}
\definecolor{awesome}{rgb}{1.0, 0.13, 0.32}
\definecolor{ballblue}{rgb}{0.13, 0.67, 0.8}
\definecolor{bittersweet}{rgb}{1.0, 0.44, 0.37}
\definecolor{blue}{rgb}{0.0, 0.0, 1.0}
\definecolor{brinkpink}{rgb}{0.98, 0.38, 0.5}
\definecolor{ballblue}{rgb}{0.13, 0.67, 0.8}
\definecolor{brightturquoise}{rgb}{0.03, 0.91, 0.87}
\definecolor{blue-green}{rgb}{0.0, 0.87, 0.87}
\definecolor{caribbeangreen}{rgb}{0.0, 0.8, 0.6}
\definecolor{cyan}{rgb}{0.0, 1.0, 1.0}
\definecolor{amber(sae/ece)}{rgb}{1.0, 0.49, 0.0}
\graphicspath{ {images/} }
\definecolor{vdarkred}{rgb}{0.6,0,0.2}

\usepackage[utf8]{inputenc}
\usepackage{lmodern}

\newtheorem{theorem}{Theorem}[section]

\definecolor{vdarkred}{rgb}{0.6,0,0.2}
\definecolor{vdarkblue}{rgb}{0,0.2,0.6}
\usepackage[pdftex, colorlinks, linkcolor=vdarkblue,citecolor=vdarkred]{hyperref}

\author{Simone Del Vecchio, J\"urg Fr\"ohlich, Alessandro Pizzo, Alessio Ranallo}

\title{Two Results in the Quantum Theory of Measurements\footnote{to appear in ``Trails in Modern Theoretical Physics. 
A Volume in Tribute of Giovanni Morchio,'' Andrea Cintio and Alessandro Michelangeli (eds.), Springer-Verlag}}

\begin{document}

\maketitle

\begin{center}
{\large Dedicated to the memory of our colleague, teacher and friend \textit{Gianni Morchio}}
\end{center}

\vspace{1em}

\begin{abstract}
Two theorems with applications to the quantum theory of measurements are stated and proven. 
The first one clarifies and amends von Neumann's Measurement Postulate used in the Copenhagen 
interpretation of quantum mechanics. The second one clarifies the relationship between ``events'' and
``measurements'' and the meaning of measurements in the $ETH$-Approach to quantum mechanics. 
\end{abstract}

\tableofcontents

\section{Introduction and summary of contents}\label{Intro}

In this paper, we present two mathematical results of relevance to the quantum theory of 
measurements,\footnote{As far as we remember, Gianni Morchio had an interest in the foundations 
of quantum mechanics; so he would probably have appreciated our results.} which we treat in a spirit 
close to the Copenhagen interpretation/heuristics of quantum mechanics (QM), as amended in \cite{FS-1, FP, FGP}.

Let $\mathfrak{E}$ be a large ensemble of physical systems identical (isomorphic) to a specific
system, $S$, of finitely many degrees of freedom to be described quantum-mechanically. We are
interested in understanding the effect of measurements of a physical quantity, $\widehat{X}$,
characteristic of $S$ for all systems in $\mathfrak{E}$. In text-book QM, one tends to invoke 
\textit{von Neumann's measurement postulate} (see \cite{JvN}) to predict properties of the resulting state,
averaged over all systems in $\mathfrak{E}$, right after a successful completion of the measurements 
of $\widehat{X}$. The standard formulation of this postulate appears 
to be afflicted with some problems, which we will discuss and attempt to clarify in the following.

We begin this paper by describing the systems $\simeq S$ we have in mind.  A physical quantity,
$\widehat{X}$, characteristic of $S$ is represented by a self-adjoint operator, $X=X^{*}$, acting 
on a separable Hlbert space, $\mathcal{H}$. An average over $\mathfrak{E}$ of states of these 
systems is called an \textit{``ensemble state''} and is given by a \textit{density matrix,} i.e., 
by a positive, trace-class operator, $\Omega$, on $\mathcal{H}$ of trace $\text{tr}\,\Omega =1$.
To mention an example, a system $S\in \mathfrak{E}$ might consist of a particle, such as an electron, 
propagating in physical space $\mathbb{E}^{3}$, $\widehat{X}$ might be a component of the spin or a 
bounded function of a component of the position- or the momentum of the particle, and
$$\mathcal{H}= L^{2}(\mathbb{R}^{3}, d^3 x) \otimes \mathbb{C}^{2s+1},$$
where $x\in \mathbb{R}^{3}$ is the position and $s$ the spin of the particle.

The purpose of this paper is to clarify what is meant by the statement that a measurement
of the quantity $\widehat{X}$ has been completed successfully.
Since we will try to follow the spirit of the Copenhagen Interpretation/heuristics of QM, where
appropriate, we will usually adopt an \textit{ensemble point of view}, emphasizing statements that 
are obtained by taking averages over all systems in the ensemble $\mathfrak{E}$. But when combined 
with results in \cite{FP, FGP}, our results have implications relevant for the theory of measurements carried
out on individual systems.

Next, we outline the contents of this paper. In Sect.~2, we recall von Neumann's measurement postulate
and point out some problems with it. We then formulate a revised version of this postulate and state the
main result proven in this paper. In Sect.~3 we sketch how measurements are described in the $ETH$- Approach to 
quantum mechanics \cite{FS-1, FP, FGP}. In Sect.~4, we present the proof of our main result. 

\section{Von Neumann's Measurement Postulate}
We imagine that the initial ensemble state when measurements of $\widehat{X}$ set in, for all systems 
in $\mathfrak{E}$, is described by a density matrix $\Omega_{in}$. In his book \cite{JvN} on the 
foundations of QM, von Neumann postulated that, when averaging over $\mathfrak{E}$, the effect 
of measuring $\widehat{X}$ for all systems belonging to $\mathfrak{E}$ amounts to replacing the 
state $\Omega_{in}$ by a certain ensemble state, $\Omega_{out}$, describing the average 
of states of systems belonging to $\mathfrak{E}$ right \textit{after} the measurements 
of $\widehat{X}$ have been completed, where $\Omega_{out}$ satisfies the following postulate.\vspace{0.15cm}\\
\textit{\underline{Von Neumann's Postulate}:}\\
\textit{Let $X=X^{*}$ be the self-adjoint operator on $\mathcal{H}$ representing the physical quantity 
$\widehat{X}$, and let
\begin{equation}\label{obs}
X= \int_{\mathbb{R}} \xi\, d\Pi(\xi)
\end{equation}
be the spectral decomposition of $X$, with $\Pi(\Delta)$ its spectral projection associated with
an arbitrary Borel set $\Delta \subset \mathbb{R}$.
The ensemble state $\Omega_{out}$ right after completion of the measurements of $\widehat{X}$ 
has the properties that}
\begin{align}\label{Neumann}
\begin{split}
&[\Omega_{out}, X] = 0, \qquad \text{and} \\
\text{tr}\big(\Omega_{in} \cdot \Pi(\Delta)\big) &= \text{tr}\big(\Omega_{out} \cdot \Pi(\Delta)\big), \,\,\forall\, 
\text{ Borel sets }\,\,\Delta  \subset \mathbb{R}\quad (Born's\,\, Rule)\quad \,\, \square
\end{split}
\end{align}

\noindent
\textit{\underline{Remark}:} We will see shortly that this formulation of von Neumann's postulate is inadequate, 
except if the operator $X$ has pure point spectrum (for which case it was originally formulated) -- but
even then it is problematic, as will become apparent in Sect.~3.

The spectral decomposition of a density matrix $\Omega$ has the form
\begin{align}\label{density matrix}
\begin{split}
\Omega&= \sum_{n=1}^{N} \omega_n \, \pi_n, \qquad1\geq \omega_1>\omega_2>\cdots >\omega_N >0,\\ 
\pi_n&=\pi_{n}^{*}, \quad \pi_n \cdot \pi_m = \delta_{nm} \pi_n\,, \quad \forall\,\,n, m =1,2,\dots, N,
\end{split}
\end{align}
for some $N\leq \infty$. The operators $\pi_n$ are disjoint orthogonal projections of finite rank (the eigenprojections
of $\Omega$), and
$$\text{tr}\big(\Omega\big) = \sum_{n=1}^{N} p_n  = 1, \quad\text{where }\,\, \,p_n = \omega_n\cdot \text{dim}\,\pi_n, \,\,\, n=1,2,\dots, N\,.$$
We set 
$$\pi_{\infty}:= \mathbf{1}- \sum_{n=1}^{N}\pi_n\,.$$
If, as in the formulation of von Neumann's postulate given in Eq. \eqref{Neumann}, the operator $X$ is assumed 
to commute with $\Omega_{out}$, then it satisfies the identity
\begin{equation}\label{specX}
X= \sum_{n=1}^{N} \pi_{n}\, X\, \pi_{n} + \pi_{\infty}\,X\, \pi_{\infty}, \quad \text{ where } \,\, \pi_{n}\,X\,\pi_{n} \,\,\text{is
of finite rank, }\, \forall\,\,n=1,2, \dots, N\,.
\end{equation}
We observe that, for every $n=1,2,\dots, N$, $\pi_{n}X\pi_{n}$ is a selfadjoint, finite-dimensional matrix; hence its spectrum 
consists of finitely many (discrete) eigenvalues. Let $\mathcal{H}^{+}$ be the subspace of $\mathcal{H}$ given
by the range of $\mathbf{1}- \pi_{\infty}$. It follows that if $X$ satisfies \eqref{Neumann} then
the operator $X\big|_{\mathcal{H}^{+}}$ has \textit{pure-point spectrum}. (Of course, if the range of $\pi_{\infty}$
is infinite-dimensional then $\pi_{\infty}\,X\, \pi_{\infty}$ may have continuous spectrum; but this is irrelevant 
for measurements of $\widehat{X}$ that result in states occupied by the systems in $\mathfrak{E}$ whose 
average is given by $\Omega_{out}$.) Thus, at best, von Neumann's postulate in the formulation 
of Eq.~\eqref{Neumann} can only be applied to measurements of physical quantities with pure-point spectrum.
However, a component of the position or of the momentum of a quantum particle propagating 
in physical space $\mathbb{E}^{3}$ has simple continuous spectrum occupying the entire real line $\mathbb{R}$. 

We conclude that Eqs.~\eqref{Neumann} \textit{cannot} be valid verbatim when physical quantities represented 
by operators with \textit{continuous spectrum} are measured, and we should find out how to modify 
them in such instances. 

\subsection{An amended form of von Neumann's Postulate}
We imagine that measurements of a physical quantity $\widehat{X}$ are carried out for all systems 
belonging to a large ensemble $\mathfrak{E}$ of systems identical to a system $S$, with the result that 
the average over $\mathfrak{E}$ of the final states of these systems after completion of the 
measurements of $\widehat{X}$ is found to be \textit{close} (but not necessarily equal) to an esnsemble state
given by a density matrix $\Omega_{out}$ with the property that
\begin{equation}\label{almostcomm}
\big|\big| \big[\Omega_{out}, X\big]\big|\big| < \varepsilon \,,
\end{equation}
for some $\varepsilon$ smaller than the error margin of the instrument used to measure $\widehat{X}$.
One may add the assumption that, for $\Omega_{out}$, \textit{Born's Rule} holds, as formulated 
in the second equation of \eqref{Neumann}. We will establish the following\vspace{0.15cm}\\
\textit{\underline{Main Result}:}
If condition \eqref{almostcomm} holds for a sufficiently small $\varepsilon\ll 1$
then one may replace $\Omega_{out}$ by a modified density matrix $\Omega_{out}'$ and $X$ 
by a modified operator $X'$,
\begin{equation}\label{pure-point}
X'= \sum_{k=1}^{K} \xi_k \,\Pi_k\,, \qquad \text{for some }\,\,K\leq \infty\,,
\end{equation}
where $\xi_1> \xi_2 >\dots >\xi_K>-\infty$ are the eigenvalues of $X'$ and $\Pi_1, \dots, \Pi_K$ the corresponding 
eigen-projections, with the properties that
\begin{enumerate}
\item[(i)]{the operator $X'$ has pure-point spectrum and is close to the operator $X$ representing
$\widehat{X}$ in the operator norm;}
\item[(ii)]{the density matrix $\Omega'_{out}$ is close to the density matrix $\Omega_{out}$ in the trace norm; and}
\item[(iii)]{the operators $X'$ and $\Omega_{out}'$ commute, i.e.,
\begin{equation}\label{comm}
\big[\Omega_{out}', X'\big]=0\,.
\end{equation}
}
\end{enumerate}
The closeness of $X'$ to $X$ and of $\Omega_{out}'$ to $\Omega_{out}$ depends on the size of the 
commutator of $X$ with $\Omega_{out}$: the smaller the norm, $\Vert [\Omega_{out}, X]\Vert$, 
of this commutator the closer are $X'$ to $X$ and $\Omega'_{out}$ to $\Omega_{out}$. The size
of $\Vert [\Omega_{out}, X]\Vert$ is thus a measure for the precision of the instrument used to measure 
$\widehat{X}$ -- the smaller this norm, the higher the precision of the instrument.

The proof of the \textit{Main Result} stated above is given in Sect.~4. At the end of the present section, 
we sketch the very easy proof in the special case where dim$(\mathcal{H}) < \infty$.\\

\textit{\underline{Remarks}:}
\begin{enumerate}
\item[(1)] {Another possible amendment of von Neumann's postulate can be formulated as follows.
We cover the spectrum, spec($X$), of the operator $X$ with small closed intervals $\Delta_k \subset \mathbb{R}, 
k= 1,2,\dots, K,$ for some $K<\infty$, with the properties that $\Delta_k \cap \Delta_{k'}$ is empty or consists
of a single point (assumed not to be an eigenvalue of $X$) whenever $k\not= k'$, and 
$\bigcup_{k=1}^{K} \Delta_k \supseteq \text{spec}(X)$. These intervals are assumed to be determined 
by properties of the instrument used to measure $\widehat{X}$. One may then assume that the 
$\mathfrak{E}$-average of the states of the systems after completion of the measurements of 
$\widehat{X}$ is given by a density matrix $\Omega_{out}$ satisfying
\begin{equation}\label{out state}
\Omega_{out}= \sum_{k=1}^{K} \Pi_k \Omega_{out}\Pi_k, \quad \text{ where }\quad \Pi_k=\Pi(\Delta_k),\, \,\,\forall 
\,\,k=1,2,\dots,K\,.
\end{equation}
The operator $X'$ is chosen to be given by
$$X'= \sum_{k=1}^{K} \xi_k\, \Pi_k,$$
where $\xi_k$ is the midpoint of the interval  $\Delta_k \subset \mathbb{R}$, for all $k$. 
Assuming that the length of all the intervals $\Delta_k$ is bounded above by $2\varepsilon$, 
we conclude that
\begin{align}\label{JvN-amended}
\big|\big|\big[\Omega_{out}, X \big]\big|\big| < \varepsilon\,, \quad
\big[\Omega_{out}, X'\big] = 0\,, \quad \text{and} \quad \Vert X-X'\Vert < \varepsilon\,.
\end{align}
This amendment of von Neumann's postulate is somewhat arbitrary and involves assumptions on what is meant 
by a measurement of a physical quantity that are more detailed than condition \eqref{almostcomm}.}
\item[(2)] {The \textit{Main Result} stated above is reminiscent of a theorem that says that if two bounded 
self-adjoint operators almost commute then there are two operators close in norm to the original ones 
that \textit{do} commute; see \cite{Lin, Hastings, Kach}.}
\item[(3)] {We conjecture that our \textit{Main Result} is a special case of the following more general 
statement: Let $\mathfrak{A}$ be a von Neumann algebra with unit $\mathbf{1}$, and let
$\omega$ be a normal state on $\mathfrak{A}$. For an operator $X\in \mathfrak{A}$, we define
a bounded linear functional on $\mathfrak{A}$ by
\begin{equation}\label{ad}
\text{ad}_X [\omega] (Y):= \omega([Y,X])\,, \quad \forall\,\, Y \in \mathfrak{A}\,.
\end{equation}
Suppose now that $\omega$ and $X$ are such that
\begin{equation}\label{upper bound}
\big| \text{ad}_X [\omega] (Y)\big| < \varepsilon \Vert Y \Vert, \quad \forall\,\, Y \in \mathfrak{A},\,\,\, \text{ for some }\,\, \varepsilon \ll 1\,.
\end{equation}
Then there exist a normal state $\omega'$ on $\mathfrak{A}$ and an operator $X' \in \mathfrak{A}$, with
$\Vert \omega' - \omega \Vert < \delta(\varepsilon)$ and $\Vert X' - X\Vert < \delta(\varepsilon)$, for 
some $\delta(\varepsilon) \searrow 0$, as $\varepsilon \searrow 0$, such that 
\begin{equation}\label{ad=0}
\text{ad}_{X'} [\omega'] = 0\,.
\end{equation}
Our \textit{Main Result} shows that this conjecture holds in the special case where $\mathfrak{A}$ is
isomorphic to the algebra of all bounded operators on a separable Hilbert space.}
\end{enumerate}
As a warm-up we prove the \textit{Main Result} in the special case of a finite-dimensional 
Hilbert space $\mathcal{H}$, which is very easy. In items (i) through (iii), one may then set 
$\Omega'_{out} = \Omega_{out}$ and only slightly modify the operator $X$, or one 
may set $X'=X$ and only slightly modify $\Omega_{out}$, and end up with \eqref{comm}. 

Let $\mathcal{H}= \mathbb{C}^M,$ with $M < \infty$. Then 
\begin{align}\label{findim}
\begin{split}
X=&\sum_{k=1}^{K} \xi_k \,\Pi_k, \qquad K\leq M\,, \quad \text{and}\\
\Omega=&\sum_{n=1}^{N} \omega_n \,\pi_n, \qquad N\leq M\,,
\end{split}
\end{align}
where $\xi_1 > \xi_2 > \dots > \xi_K> -\infty$ are the eigenvalues of $X$ and $\Pi_1, \Pi_2, \dots, \Pi_K$ 
are the corresponding eigen-projections, and $\omega_1 > \omega_2 >\dots > \omega_n > 0$ are the 
non-zero eigenvalues of $\Omega$, with $\pi_1, \pi_2, \dots, \pi_N$ the corresponding eigen-projections. 
We define $\pi_{N+1}:= \mathbf{1}- \sum_{n=1}^{N} \pi_n$, and
\begin{equation}\label{gap}
\gamma_\Omega := \underset{1\leq n\leq N}{\text{min}} \big(\omega_n - \omega_{n+1}\big) >0, \qquad 
\text{with }\,\,  \omega_{N+1}:=0\,,
\end{equation}
to be the smallest gap between distinct eigenvalues of $\Omega$. Let us assume that
\begin{equation}\label{smallcomm}
\big|\big| \big[X, \Omega \big]\big|\big| \leq \varepsilon, \quad \text{for some }\, \varepsilon \ll \gamma_\Omega\,.
\end{equation}
We define an operator $X'$ by setting
\begin{equation}\label{defX'}
X':= \sum_{n=1}^{N+1} \pi_n \, X\, \pi_n\,.
\end{equation}
Obviously $X'$ commutes with $\Omega$, and we claim that 
\begin{equation}\label{closeness-1}
\Vert X' - X\Vert < \text{const.}\, \varepsilon\,.
\end{equation}
\textit{Proof of \eqref{closeness-1}}. Clearly
\begin{equation}\label{id-1}
X= \sum_{n, n' = 1,2, \dots, N+1} \pi_n \,X\, \pi_{n'}\,.
\end{equation}
By \eqref{smallcomm} we have that
\begin{equation}\label{id-2}
\Vert [\pi_n \,X\, \pi_{n'}, \Omega] \Vert = \Vert \pi_n [X,\Omega] \pi_{n'}\Vert \leq \varepsilon, \quad \forall \,\, n, n' \,.
\end{equation}
Plainly $[\pi_n \,X\, \pi_{n}, \Omega]= 0\,, \,\, \forall\,\, n=1,2, \dots, N+1$. If $n \not= n'$ then
$$[\pi_n \, X\, \pi_{n'}, \Omega] = (\omega_{n'} - \omega_n)\, \pi_n \, X\, \pi_{n'}\,.$$
By Eqs.~\eqref{gap} and \eqref{id-2}, we have that
$$\Vert \pi_n \, X\, \pi_{n'} \Vert \leq \gamma_{\Omega}^{-1}\,\varepsilon\,, \quad \text{for }\,\, n\not= n'\,.$$
Thus, using \eqref{id-1} we find that
\begin{equation}\label{normbound}
\Vert X- X' \Vert \leq (N+1)N\,\gamma_{\Omega}^{-1}\, \varepsilon < M^{2} \gamma_{\Omega}^{-1} \varepsilon\,,
\end{equation}
as claimed in \eqref{closeness-1}.

In the calculations just shown we can obviously exchange the roles of $X$ and $\Omega$. We set
$$\gamma_X := \underset{1\leq k< K}{\text{min}} \big(\xi_k - \xi_{k+1}\big) >0\,,$$
and we then replace the density matrix $\Omega$ by
$$\Omega':= \sum_{k=1}^{K} \Pi_k\, \Omega\, \Pi_k\,.$$
Clearly $\Omega'$ is a non-negative operator, and tr$(\Omega') = 1$, because $\sum_{k=1}^{K} \Pi_k = \mathbf{1}$; 
i.e., $\Omega'$ is a density matrix; and it obviously commutes with $X$.
Repeating the arguments shown above, we find that
\begin{equation}\label{tracebound}
\text{tr}( |\Omega - \Omega' |) \leq M\,(K-1)K \,\gamma_{X}^{-1}\, \varepsilon < M^{3}\, \gamma_{X}^{-1} \varepsilon\,.
\end{equation}
Of course, the problem with the estimates in \eqref{normbound} and \eqref{tracebound} is the dependence
of the right sides on the dimension, $M$, of the Hilbert space $\mathcal{H}$. This problem is addressed in
Sect.~4, where we state a result that is \textit{uniform} in the dimension of the Hilbert space, but at the
price that we have to slightly modify both, $X$ and $\Omega$. This result enables one to modify von Neumann's measurement postulate so as to avoid the shortcomings of the original version, as indicated above.\\

\section{The Description of Measurements in the $ETH$-Approach to QM}
In this section we sketch how measurements can be described in the formulation of QM proposed in
\cite{FS-1, FP, FGP} under the name of ``$ETH$-Approach to QM'' (assuming some familiarity with
these papers).

We begin with the obvious observation that a successful measurement of a physical quantity $\widehat{X}$ 
characteristic of a system $S$ (belonging to an exnsemble $\mathfrak{E}$) results in an \textit{event}, 
namely the event that $\widehat{X}$ takes a -- possibly somewhat imprecise -- \text{value} belonging 
to some small interval contained in the real line whose length depends on the accuracy of the 
instrument used to measure $\widehat{X}$. 
To understand the significance of this statement it is necessary to clarify what, in the $ETH$-Approach
to QM, is meant by an \textit{``event''}. We recall the definition proposed in \cite{FP, FGP}. 
Abstractly, a \textit{``potential event''}, $\mathfrak{e}$, associated with a physical system $S\in \mathfrak{E}$ 
is a partition of unity, $\mathfrak{e}=\big\{\pi_n\big\}_{n=1}^{\infty}$, by orthogonal projections satisfying
\begin{equation}\label{partition}
\pi_n = \pi_{n}^{*}, \quad \pi_n \cdot \pi_{n'} = \delta_{n n'}\,\pi_n, \,\, \forall\,\, n, n'=1,2,\dots, \quad 
\sum_{n=1}^{\infty} \pi_n = \mathbf{1}\,.
\end{equation}
An operator $X$ representing a physical quantity $\widehat{X}$ characteristic of a system $S\in \mathfrak{E}$
at some time $\geq t$ and the projections $\pi\in \mathfrak{e}$ of an arbitrary potential event 
$\mathfrak{e}$ that may occur in $S$ at a time $\geq t$ are supposed to belong to some algebra 
$\mathfrak{A} = \mathcal{E}_{\geq t}$, which, in general, depends \textit{non-trivially} on time $t$. 
For systems, $S$, with finitely many degrees of freedom, $\mathfrak{A}$ is the algebra, $B(\mathcal{H})$, 
of all bounded operators on a separable Hilbert space $\mathcal{H}$ and is independent of $t$. 
But, for systems with infinitely many degrees of freedom, including those describing the quantized 
electromagnetic field,\footnote{the only systems for which (in our opinion) the ``measurement problem'' has a 
satisfactory solution} the time-dependence of $\mathfrak{A}=\mathcal{E}_{\geq t}$ tends to be 
\textit{non-trivial}, and $\mathfrak{A}$ is a more exotic (type-III$_{1}$) algebra. 
Our analysis in this section does not require any specific assumptions on 
$\mathfrak{A}$. (It is only assumed that the algebra $\mathfrak{A}$ is weakly closed, 
i.e., that it is a von Neumann algebra; but it need not and usually will \textit{not} be isomorphic to $B(\mathcal{H})$.)
States at time $t$ are states on $\mathfrak{A}=\mathcal{E}_{\geq t}$ (i.e., positive, 
normalized linear functionals on $\mathcal{E}_{\geq t}$). They are denoted  by lower-case Greek 
letters, $\omega, \dots$. 

In the following discussion we fix a time $t$ and suppress explicit reference to time-dependence wherever 
possible. We suppose that a state, $\omega$, on $\mathfrak{A}$ is an \textit{ensemble state,} i.e., that it has the 
meaning of being an average over the ensemble $\mathfrak{E}$ of states of individual systems, 
all $\simeq S$. If a potential event $\mathfrak{e}=\big\{\pi_n\big\}_{n=1}^{\infty}\subset \mathfrak{A}$ is \textit{actualizing} 
(i.e., is observed to happen) at some time $t$ then, according to the $ETH$- Approach 
to QM, the state $\omega= \omega_t$ has the property that
\begin{equation}\label{actuality}
\omega(X)= \sum_{\pi \in \mathfrak{e}} \omega( \pi\cdot X\cdot \pi)\,,\qquad \forall\,\,\, X\in \mathfrak{A}\,,
\end{equation}
i.e., $\omega$ is a convex combination of states in the images of the projections $\pi\in \mathfrak{e}$.
Potential events actualizing at some time are called \textit{``actualities''}. (For a more precise 
characterization of actualities, see, e.g., \cite{FGP}.) If $\mathfrak{A}=B(\mathcal{H})$ then
$$\omega(X)= \text{tr}\big(\Omega\cdot X\big)\,, \quad \forall \,\,\, X\in \mathfrak{A}\,,$$
for some density matrix $\Omega$ on $\mathcal{H}$, and the projections $\pi$ belonging to
the event $\mathfrak{e}$ that actualizes, given the state $\omega$,
are the spectral projections of the density matrix $\Omega$.

If $\mathfrak{e}$ is an event actualizing at some time $t$ then the state at time $t$ of an \textit{individual} 
system in the ensemble $\mathfrak{E}$ is expected to belong to the image of a projection $\pi \in \mathfrak{e}$,
with a probabilty, $prob_{\omega}(\pi)$, given by \textit{Born's Rule}, namely 
$$prob_{\omega}(\pi)= \omega(\pi)\,,$$
where $\omega$ is the ensemble state at time $t$.

We are interested in characterizing actualities $\mathfrak{e}=\big\{\pi_n\big\}_{n=1}^{N} \subset \mathfrak{A},\,
N\leq \infty,$ that can be interpreted as corresponding to the completion of the measurement of a 
certain physical quantity $\widehat{X}$. 
We thus consider a state $\omega$ satisfying Eq.~\eqref{actuality}. Given a non-negative number 
$\varepsilon \ll 1$, there exists an integer $N_0 < \infty$ such that 
\begin{equation}\label{tail}
\sum_{n=1}^{N_0-1} \omega(\pi_n) > 1- \varepsilon\,, \,\,i.e., \quad \omega\big(\pi^{(N_0)}\big) < \varepsilon\,, 
\quad\text{where} \quad \pi^{(N_0)}:= \sum_{n=N_0}^{N} \pi_n\,. 
\end{equation}
It is then very unlikely that an individual system in $\mathfrak{E}$ is found to occupy a state 
in the range of a projection $\pi \leq \pi^{(N_0)}$. If $\mathfrak{e}$ is the potential event 
actualizing at a certain time $t$ and $\omega$ is the ensemble state at time $t$ satisfying \eqref{actuality} 
then the slightly coarser event $\mathfrak{e}_0:= \big\{\pi_1, \pi_2, \dots, \pi_{N_0 -1}, \pi^{(N_0)}\big\}$ 
can be viewed to be an actuality at time $t$, too. To avoid irrelevant complications, we henceforth replace 
$\mathfrak{e}$ by $\mathfrak{e}_0$ throughout the following discussion, and we simplify our notations by 
writing $\mathfrak{e}$, instead of $\mathfrak{e}_0$, and $\pi_{N_0}$, instead of $\pi^{(N_0)}$, 
with $N_0 < \infty$.

We assume that the operator $X$ representing the physical quanitiy $\widehat{X}$ has the form
\begin{equation}\label{finite spec}
X=\sum_{k=1}^{K} \xi_k \Pi_k, \quad \text{ for some }\,\,\,K< \infty\,,
\end{equation}
where the real numbers $\xi_k$ are the eigenvalues of $X$ and the operators $\Pi_k$ are the 
corresponding eigen-projections, $k=1,2,\dots,K.$
(We should mention that the projections $\Pi_k$ may be given by $\Pi_k= \Pi(\Delta_k)$, where the sets 
$\Delta_k$ are intervals of the real line of length $< 2\varepsilon$ whose union covers spec$(X)$, 
and $\xi_k$ may be (e.g.) the midpoint of the interval $\Delta_k$, for all $k$, as discussed in Remark (1) 
of Subsect.~2.1.)

If the actuality $\mathfrak{e}$ can be interpreted to correspond to the likely completion of a measurement 
of $\widehat{X}$, with an accuracy measured by $\varepsilon$, then there must exist a decomposition of 
$\big\{1,2, \dots, N_0\big\}$ into disjoint subsets $\mathcal{I}_k,\, k=1,2, \dots K,$ such that
\begin{align}\label{meas}
\begin{split}
\Vert [\pi_n, \Pi_k] \Vert& < \mathcal{O}\big(N_{0}^{-2} \varepsilon\big), \quad \forall \,\, n\leq N_0, \quad \text{ and }\\
\sum_{n \not\in \mathcal{I}_k, n<N_0} \Vert\pi_n\, \Pi_k\, &\pi_n \Vert < \mathcal{O}(\varepsilon), \qquad \forall \,\, k=1,2,\dots K\,.
\end{split}
\end{align}
The second equation tells us that if a system is found in a state in the range of a projection 
$\pi_n, n\not\in \mathcal{I}_k, n<N_0,$ then the quantity $\widehat{X}$ is very unlikely to have the 
measured value $\xi_k$. By \eqref{tail}, if the ensemble state is given by $\omega$ 
then it is very unlikely that an individual system in $\mathfrak{E}$ is found in a state belonging 
to the range of the projection $\pi_{N_0}$.

Since $\sum_{n=1}^{N_0} \pi_n = \mathbf{1}$, one obviously has that
$$X= \sum_{n, n' = 1,2,\dots, N_0} \pi_n \, X\, \pi_{n'}\,.$$
Since $\pi_n\cdot \pi_{n'} = 0,$ for $n\not= n'$, the first inequality in \eqref{meas} then implies 
that the operator $X$ is approximated in norm by
\begin{equation}\label{first approx}
X' := \sum_{n=1}^{N_0} \pi_n\,X\, \pi_n\,,
\end{equation}
up to an error of $\mathcal{O}(\varepsilon)$; and \eqref{tail} tells us that the Born probability of picking 
up a correction in determining the outcome of the measurement of $\widehat{X}$ that is due to the operator 
$\pi_{N_0}\,X\, \pi_{N_0}$ is bounded by $\mathcal{O}(\varepsilon)$, hence very small.
One may then wonder whether the actuality $\mathfrak{e}$ could occur as the result of a measurement 
of a \textit{slightly different} physical quantity $\simeq \widehat{X}$. 

The second inequality in \eqref{meas} implies that $X'$ is well approximated by the operator
\begin{equation}\label{third approx}
X'':= \sum_{k=1}^{K} \sum_{n\in \mathcal{I}_k} \xi_k \, \pi_n\, \Pi_k \, \pi_n + \pi_{N_0}\, X\, \pi_{N_0}
\end{equation}
with
\begin{equation}\label{small error}
\Vert X'' - X' \Vert < \mathcal{O}(K\,\varepsilon)\,.
\end{equation}
Next, we note that the first inequality in \eqref{meas} implies that
$$\big|\big| \big(\pi_n\, \Pi_k\, \pi_n\big)^{2} - \pi_n\, \Pi_k\, \pi_n \big|\big| < \mathcal{O}\big(N_{0}^{-2} \varepsilon\big)\,.$$
This estimate enables us to apply the following\\

\noindent
\textit{\underline{Lemma}.}
\textit{Let $P$ be a self-adjoint operator in a von Neumann algebra $\mathfrak{A}$, and let $\delta < \frac{1}{2}$.
If $\Vert P^{2}-P\Vert < \delta$ then there exists an orthogonal projection $\widehat{P}\in \mathfrak{A}$ 
whose image belongs to the range of $P$ such that}
$$\Vert \widehat{P} - P \Vert < \delta\,.$$
See Lemma 8 and Appendix C of \cite{FS-2}.  This lemma implies that if $N_{0}^{-2} \varepsilon$ is small enough
then there exists an orthogonal projection $\pi_{k, n}$ with the property that the image of $\pi_{k, n}$ 
is contained in or equal to the image of $\pi_n$ and such that
$$\Vert \pi_{k,n} - \pi_n\, \Pi_k \, \pi_n \Vert < \mathcal{O}\big(N_{0}^{-2} \varepsilon\big)\,.$$
We define
\begin{equation}\label{approx quantity}
X''':= \sum_{k=1}^{K} \xi_k \big(\sum_{n \in \mathcal{I}_k} \pi_{k,n}\big) + \pi_{N_0}\, X\, \pi_{N_0}, \quad \text{and }
\quad X_{fin}:= X'''- \pi_{N_0}\, X\, \pi_{N_0}\,.
\end{equation}

We are ready to state a result in the theory of measurements, according to the $ETH$-Approach to QM.
\begin{theorem}
We assume that the bounds in \eqref{tail} and \eqref{meas} hold for some $\varepsilon \ll 1$. Then we have that
\begin{enumerate}
\item[(i)]{the Born probability of finding an individual system in the ensemble $\mathfrak{E}$ in a state
that belongs to the range of the projection $\pi_{N_0}= \pi^{(N_0)}$ is bounded above by $\varepsilon$;}
\item[(ii)]{the operator $X'''$ defined in \eqref{approx quantity} is reduced by the projections 
$\pi_{N_0}$ and $\mathbf{1} - \pi_{N_0}$; }
\item[(iii)]{the norm of $X'''- X$ is bounded by $$\Vert X'''-X\Vert < \mathcal{O}(\varepsilon),$$
i.e., the physical quantity $\widehat{X}$ is well approximated by a slightly modified physical quantity 
represented by the operator $X'''$;}
\item[(iv)]{the eigenvalues of $X_{fin}= X'''- \pi_{N_0}\,X\,\pi_{N_0}$ are contained in or equal 
to the spectrum, $\big\{\xi_k\big\}_{k=1}^{K},$ of the operator $X$ and the eigen-projection of $X_{fin}$ 
corresponding to $\xi_k$ is given by the projection 
$\sum_{n\in \mathcal{I}_k} \pi_{k,n} \,\big($which is dominated by the projection 
$\sum_{n\in \mathcal{I}_k} \pi_n\big)$, for $k=1,2,\dots, K$; and
$$\big[X_{fin}, \pi_n\big] =0, \quad \forall\,\,\, \pi_n \in \mathfrak{e}\,.\qquad \square$$}
\end{enumerate}
\end{theorem}

We conclude that, under the hypotheses of Theorem 3.1, one may interpret the actualization of the 
event $\mathfrak{e}$ as being accompanied by the completion of a measurement of a physical 
quantity $\widehat{X}''' \approx \widehat{X}$, where $\widehat{X}'''$ is represented by an operator $X'''$
that is a tiny modification of the operator $X$ representing $\widehat{X}$.

In this section, we have not tried to optmize our results; we have attempted to outline the basic ideas
of how measurements can be interpreted in the $ETH$-Approach described in \cite{FS-1, FP, FGP}.

\section{Proof of the Main Result}
In this section we prove the \textit{Main Result} announced in Sect.~2. 
We consider a density matrix $\Omega$ on a separable Hilbert space $\mathcal{H}$ with spectral 
decomposition
\begin{equation}\label{spect dec}
\Omega= \sum_{n=1}^{\infty} \omega_n\, \pi_n\,,  \qquad \omega_1 >\omega_2 > \cdots\,.
\end{equation}
as in Eq.~\eqref{density matrix} of Sect.~2. We define $p_n:= \omega_{n} \cdot \text{dim} \pi_n, \, n=1,2, \dots$ 
Given a positive number $\varepsilon \ll 1$, we define $\Delta_{\varepsilon}$ by
\begin{equation}\label{tail-2}
\Delta_{\varepsilon}:= \sum_{n\,:\, \omega_n \leq \varepsilon^{1/4}} p_n\,.\end{equation}
Clearly, $\Delta_{\varepsilon} \searrow 0$, as $\varepsilon \searrow 0$.
The \textit{Main Result} is a consequence of the following theorem.
\begin{theorem}
Let $\Omega$ and $\Delta_{\varepsilon}$ be as in \eqref{spect dec} and \eqref{tail-2}, respectively,
and let $X$ be a self-adjoint operator on $\mathcal{H}$, with $\Vert X \Vert \leq 1$. We assume that
\begin{equation}\label{tiny comm}
\big|\big|\big[\Omega, X \big]\big|\big| \leq \varepsilon\,.
\end{equation}
Then, for sufficiently small values of $\varepsilon$ and $\Delta_{\varepsilon}$, there exist a density
matrix $\Omega'$ and a self-adjoint operator $X'$ such that
\begin{equation}\label{bounds}
\Vert X-X' \Vert \leq \varepsilon^{1/4}, \quad \text{and} \quad \text{tr}\big|\Omega-\Omega' \big|
\leq 2\Delta_{\varepsilon}+\mathcal{O}(\varepsilon^{1/4})\,.
\end{equation}
\end{theorem}
\textbf{Proof}.\\
As announced in the theorem, our goal is to construct a density matrix $\Omega'$ close to $\Omega$ 
in the trace norm and a self-adjoint operator $X'$ close to $X$ in the operator norm such that 
$[\Omega', X'] = 0$. We begin with the construction of $\Omega'$.

In the following it is convenient to rewrite the spectral decomposition of $\Omega$ as follows:
\begin{equation}\label{eigenvectors}
\Omega= \sum_{j=1}^{\infty} \omega_j \,|u_j\big>\,\big<u_j|\,, \qquad \omega_1\ge\omega_2\geq\cdots\geq 0, \quad
\sum_{j=1}^{\infty}\omega_j =1\,,
\end{equation}
where $\big\{u_j \big\}_{j=1}^{\infty}$ is an orthonormal system of eigenvectors of $\Omega$, and
$|u_j\big>\,\big<u_j|$ is the orthogonal projection onto $u_j$, for all $j$. Then assumption \eqref{tiny comm}
implies that
\begin{equation}\label{bound-2}
\Vert [\Omega, X] u_i \Vert^{2}=\sum_{j=1}^{\infty} (\omega_i - \omega_j)^{2} |\big<u_i, X u_j\big>|^{2} \leq \varepsilon^{2},
\quad \forall i\,.
\end{equation}
In the following steps, we construct a positive trace-class operator 
$\widetilde{\Omega}\leq \Omega,$ (hence $\text{tr}\,\widetilde{\Omega} \leq 1$).
\begin{enumerate}
\item[1)] {We preserve the eigenvectors of the density operator $\Omega$, but - where necessary - modify the 
corresponding eigenvalues in such a way that the spectrum of the modified operator $\widetilde{\Omega}$ 
consists of (possibly degenerate) eigenvalues separated by gaps of specified size. To begin with we choose 
two exponents, $\delta$ and $\beta$ (later set equal to $1/4$ and $3/4$, respectively),  with  
\begin{equation}
0<\delta <\beta<1\quad \text{and}\quad \beta>2\delta\,,
\end{equation}
and we modify the spectrum of $\widetilde{\Omega}$ in such a way that the gaps between the 
non-coinciding modified eigenvalues, i.e., between the distinct eigenvalues of $\widetilde{\Omega}$, 
will be larger than $\varepsilon^{\beta}$.

1-i) We observe that, since $\Omega\geq 0, \text{with }\, \text{tr}\,\Omega =1$,  the dimension of the direct 
sum of the eigenspaces of $\Omega$ corresponding to eigenvalues larger than or equal to  
$\varepsilon^{\delta}$ is  bounded above by $O(1/\varepsilon^{\delta})$.

1-ii) Next, we define $\omega_{i_1}$ to be the smallest eigenvalue of $\Omega$ of order
$\varepsilon^{\delta}$ with the property that its separation from the previous (next larger) 
eigenvalue is bounded below by $\varepsilon^{\beta}$. It is not assumed that an eigenvalue
with the properties of $\omega_{i_1}$ exists.

\noindent
But if such an eigenvalue $\omega_{i_1}$ exists then we denote by $(\omega_{i_1})_{-}$ its precursor.
By construction, we have that 
$(\omega_{i_1})_{-}\leq \mathcal{O}(\varepsilon^{\delta}+\varepsilon^{\beta-\delta})$, 
because there are at most $\mathcal{O}(\varepsilon^{-\delta})$ eigenvalues separated by
gaps bounded by $\leq \varepsilon^{\beta}$ in between $\omega_{i_1}$ and $(\omega_{i_1})_{-}$, 
as follows from 1-i). 

\noindent
We define 
\begin{itemize}
\item{an interval $I_0$ by $I_0:=[0,(\omega_{i_1})_{-}]\,,$ }
\item{and a subspace $\mathcal{H}_{0} \subset \mathcal{H}$ as the direct sum of the 
eigenspaces of $\Omega$ corresponding to eigenvalues contained in the interval $I_0$.}
\end{itemize}
If an eigenvalue with the properties of $\omega_{i_1}$ does not exists then we conclude 
that the largest eigenvalue, $\omega_{max}$, of $\Omega$ must be smaller than
$\mathcal{O}(\varepsilon^{\delta}+\varepsilon^{\beta-\delta})$. 
In this case, we define $I_0:=[0,\omega_{max}]$.

\noindent
 We define $\widetilde{\Omega}$ to vanish on the subspace $\mathcal{H}_0$. 

1-iii) We next assume that  $I_0\neq [0,\omega_{max}]$, i.e., that an eigenvalue with the properties of
$\omega_{i_1}$ exists. Then we consider the smallest eigenvalue of $\Omega$ larger than 
$\omega_{i_1}$ with the property that its separation from the previous eigenvalue is larger 
than $\varepsilon^{\beta}$. 

\noindent
If such an eigenvalue exists we denote it by $\omega_{i_2}$ and its precursor by $(\omega_{i_2})_{-},$
and we then have that $(\omega_{i_2})_{-} \leq \mathcal{O}(\omega_{i_1}+\varepsilon^{\beta-\delta})$. 
We also define 
\begin{itemize}
\item
$I_1:=[\omega_{i_1}\,,\,(\omega_{i_2})_{-}]$;
\item
$n_1 := $ number of eigenvalues (with multiplicity) of $\Omega$ contained in $I_1$; 
\item
$\mathcal{H}_1 :=$ direct sum of the corresponding eigenspaces
(notice that $\text{dim} \mathcal{H}_1=n_1$).
\end{itemize}
If an eigenvalue with the properties of $\omega_{i_2}$ does not exists we conclude that the largest 
eigenvalue, $\omega_{max}$, of $\Omega$ is smaller than $\mathcal{O}(\omega_{i_1}+\varepsilon^{\beta-\delta}),$ 
and we define $I_1:=[\omega_{i_1},\omega_{max}]$

\noindent
On the subspace $\mathcal{H}_1$  we define 
$$\widetilde{\Omega}\big|_{\mathcal{H}_1}:=\widetilde{\omega}_{1}\cdot \mathbf{1}\big|_{\mathcal{H}_1},$$ 
where $\widetilde{\omega}_{1}:=\omega_{i_1}$.

1-iv) We iterate these arguments: If $I_{m-1}\neq [\omega_{i_{m-1}}\,,\,\omega_{max}]$, then, starting from 
$\omega_{i_m}$, we consider the eigenvalue of $\Omega$ with the property that its separation from the 
previous one is bounded below by $\varepsilon^{\beta}$.

\noindent
If an eigenvalue of $\Omega$ with these properties exists we denote it by $\omega_{i_{m+1}}$ and 
the previous one by $(\omega_{i_{m+1}})_{-}\,\big(\leq \mathcal{O}(\omega_{i_m}+\varepsilon^{\beta-\delta})\big)$. 
We also define 
\begin{itemize}
\item
$I_m:=[\omega_{i_m}\,,\,(\omega_{i_{m+1}})_{-}]$;
\item 
$\mathcal{H}_m:=$ direct sum of eigenspaces of $\Omega$ corresponding to eigenvalues 
contained in the interval $I_m$; and $n_m:= \text{dim}\mathcal{H}_m$.
\end{itemize}
If this eigenvalue does not exists we conclude that the largest eigenvalue, $\omega_{max}$, of $\Omega$ is 
bounded above by $\mathcal{O}(\omega_{i_m} + \varepsilon^{\beta-\delta})$, and we define 
$I_m:=[\omega_{i_m},\omega_{max}]$
\\

\noindent
 On the subspace $\mathcal{H}_m$ we define the operator $\widetilde{\Omega}$ by 
 $\widetilde{\Omega}\big|_{\mathcal{H}_m}:=\widetilde{\omega}_m\cdot \mathbf{1}\big|_{\mathcal{H}_m}$, 
 where $\widetilde{\omega}_m:=\omega_{i_m}$.

 1-v)
 The construction described above must necessarily stop at some step $\overline{m}\geq 0$, 
 because  $\Omega$ is trace-class and $\varepsilon^{\beta}>0$.
 The spectrum of the operator $\widetilde{\Omega}$ constructed above consists of the points
 \begin{equation}
 \{\,\omega_{i_0}:=0\,,\,\omega_{i_1}\,,\,\dots\,,\, \omega_{i_{\overline{m}}}\}\,.
 \end{equation}

1-vi) We note that $\widetilde{\Omega}$ has been defined as the operator whose eigenspaces 
are the subspaces $\mathcal{H}_m$ and the corresponding eigenvalues are given by $\widetilde{\omega}_m$. 
(To avoid possible confusion we stress that the eigenvalues $\widetilde{\omega}_m$ of 
$\tilde{\Omega}$ are \textit{increasing} in $m$ whereas the eigenvalues $\omega_i$ of $\Omega$ 
are \textit{decreasing} in $i$.) The operator $\widetilde{\Omega}$ enjoys the property
 \begin{equation}
 \text{tr}|\Omega-\widetilde{\Omega}|\leq   o(1)+\mathcal{O}\big(\varepsilon^{\beta-\delta}(n_1+\dots+n_{\overline{m}})\big) \leq o(1)+O(\varepsilon^{\beta-2\delta})\,,
\end{equation}
which holds, because
\begin{equation}
0< \sum_{\omega_i\leq \varepsilon^{\delta}} \omega_i \leq o(1)\,;
\end{equation}
(recall that $\Omega$ is trace-class and that, in \eqref{tail-2}, we have noticed that
$\underset{i\,:\,\omega_i\leq \varepsilon^{1/4}}{\sum} \omega_i =: \Delta_{\varepsilon}\ll1$). Moreover, 
we use the facts that any eigenvalue of $\Omega$ corresponding to an eigenvector in $\mathcal{H}_{m}$ 
is included in the interval $[\tilde{\omega}_m\,,\,\tilde{\omega}_m+\epsilon^{\beta-\delta}]$, by construction 
of $\mathcal{H}_{m}$, and that
\begin{equation}
n_1+\dots+n_{\overline{m}}\leq O(1/\varepsilon^{\delta})\,,
 \end{equation}
as shown in 1-i).}

\item[2)] {Next, we modify the operator $X$. The modified operator is denoted by $X'$ and is defined
by its matrix elements in the basis, $\big\{u_j\big\}_{j=1}^{\infty}$, of eigenvectors of $\Omega$, which 
are given by
\begin{equation}
(X')_{i\,,\,j}:=\langle u_i\,,\,X\,u_j \rangle\,,
\end{equation}
provided that $u_i$ and $u_j$ belong to the same subspace $\mathcal{H}_p,\, p\leq \overline{m}$, 
and
\begin{equation}
(X')_{i\,,\,j}:=0\,,
\end{equation}
if $u_i$ and $u_j$ belong to \textit{different} eigenspaces, $\mathcal{H}_p$, $\mathcal{H}_{p'}$, 
of $\widetilde{\Omega}$.

\noindent
We thus have by construction that
\begin{equation}
[\,\widetilde{\Omega}\,,\,X'\,]=0\,.
\end{equation}
Next, we show that $\|X-X'\|=o(1)$. This follows from 
\begin{equation}
\sup_{i} \sum_{j=1}^{\infty} |\langle u_j\,,\, (X-X')\, u_i\rangle |^2=\varepsilon^{2(1-\beta)}\,,
\end{equation}
where the summands are non-zero only if  $u_i$ and $u_j$ belong to different eigenspaces 
$\mathcal{H}_{p_i}$, $\mathcal{H}_{p_j}$ of $\widetilde{\Omega}$, so that
\begin{equation}
\sum_{j=1}^{\infty} |\langle u_j ,\, (X-X')\, u_i\rangle |^2=\sum_{j\,:\, u_j\in \mathcal{H}_{p_j}\,,\, p_j\not=p_i}
 |\langle u_j ,\, X\, u_i\rangle |^2\,.
\end{equation}
But if $p_j\not= p_i$ then $|\omega_i-\omega_j|\geq \varepsilon^{\beta},$ where $\omega_i$ and 
$\omega_j$ are the eigenvalues of $\Omega$ on the vectors $u_j\in \mathcal{H}_{p_j}$ 
and $u_i\in \mathcal{H}_{p_i}$, respectively. Next, we exploit the bound assumed in 
\eqref{tiny comm}, namely
\begin{align}
\varepsilon^2 \geq &\, \|\,[\,\Omega\,,\,X\,]\,u_i\,\|^2\\
=& \sum_{j=1}^{\infty}\,(\omega_i-\omega_j)^2|\langle u_i\,,\,X\,u_j\,\rangle|^2\\
\geq &\, \varepsilon^{2\beta}\sum_{j\,:\, u_j\in \mathcal{H}_{p_j}\,,\, p_j\not= p_i}\,|\langle u_i\,,\,X\,u_j\,\rangle|^2
\end{align}

\noindent
To conclude the proof of the theorem, we normalize $\widetilde{\Omega}$ by dividing by its trace,
defining $\Omega':=\frac{\widetilde{\Omega}}{\text{tr}\,\widetilde{\Omega}}$. Setting 
$\delta=\frac{1}{4}$ and $\beta=\frac{3}{4}$, and using that
  $$ \text{tr}|\Omega-\widetilde{\Omega}|\leq \Delta_{\varepsilon} + \mathcal{O}(\varepsilon^{\frac{1}{4}})\,,$$ 
  we conclude that $$ \text{tr}|\Omega - \Omega'|\leq 2\,\Delta_{\varepsilon} + \mathcal{O}(\varepsilon^{\frac{1}{4}})\,.$$
  \hspace{14cm}$\square$
  }
\end{enumerate}

\begin{center}
-----
\end{center}

\bigskip

\noindent
Simone Del Vecchio, Dipartimento di Matematica, Università degli Studi di Bari, Italy, \\\href{mailto:simone.delvecchio@uniba.it}{simone.delvecchio@uniba.it}\\
J\"urg Fr\"ohlich, ETH Z\"urich, Institute for Theoretical Physics, \\ \href{mailto:juerg@phys.ethz.ch}{juerg@phys.ethz.ch}\\
Alessandro Pizzo, Dipartimento di Matematica, Università di Roma “Tor Vergata", Italy, \\ \href{mailto: pizzo@mat.uniroma2.it}{pizzo@mat.uniroma2.it}\\
Alessio Ranallo, Dipartimento di Matematica, Università di Roma “Tor Vergata", Italy, \\ \href{mailto: 
ranallo@mat.uniroma2.it}{ranallo@mat.uniroma2.it}
\end{document}